\documentclass{Interspeech}
\usepackage{todonotes}
\usepackage{xcolor}
\usepackage{dblfloatfix}
\usepackage{doi}
\usepackage{graphicx}
\usepackage{cite}
\usepackage[utf8]{inputenc}
\usepackage[T1]{fontenc}

\definecolor{arytenoid-cartilage}{rgb}{0.54, 0.17, 0.89} 
\definecolor{epiglottis}{rgb}{0.25, 0.88, 0.82} 
\definecolor{lower-lip}{rgb}{0.00, 1.00, 0.00} 
\definecolor{pharyngeal-wall}{rgb}{0.85, 0.66, 0.13} 
\definecolor{soft-palate-midline}{rgb}{0.12, 0.56, 1.00} 
\definecolor{tongue}{rgb}{1.00, 0.55, 0.00} 
\definecolor{upper-lip}{rgb}{1.00, 0.00, 1.00} 
\definecolor{vocal-folds}{rgb}{1.00, 0.41, 0.71} 
\interspeechcameraready 

\title{Reconstruction of the Complete Vocal Tract Contour Through Acoustic to Articulatory Inversion Using Real-Time MRI Data}

\author[affiliation={1}]{Sofiane}{Azzouz}
\author[affiliation={2}]{Pierre-André}{Vuissoz}
\author[affiliation={1}]{Yves}{Laprie}

\affiliation{Université de Lorraine}{CNRS, Inria}{F-54000 Nancy, France}
\affiliation{Université de Lorraine}{Inserm, IADI U1254}{-54000 Nancy, France}
\email{sofiane.azzouz@loria.fr, pa.vuissoz@chru-nancy.fr, yves.laprie@loria.fr}
\keywords{Acoustic to articulatory inversion, speech production, rt-MRI}

\usepackage{todonotes}

\newif\ifshowtodos
\showtodosfalse

\begin{document}

\maketitle


\begin{abstract}
Acoustic to articulatory inversion has often been limited to a small part of the vocal tract because the data are generally EMA (ElectroMagnetic Articulography) data requiring sensors to be glued to easily accessible articulators. The presented acoustic to articulation model focuses on the inversion of the entire vocal tract from the glottis, the complete tongue, the velum, to the lips. It relies on a realtime dynamic MRI database of more than 3 hours of speech. The data are the denoised speech signal and the automatically segmented articulator contours. Several bidirectional LSTM-based approaches have been used, either inverting each articulator individually or inverting all articulators simultaneously. To our knowledge, this is the first complete inversion of the vocal tract. The average RMSE precision on the test set is 1.65 mm to be compared with the pixel size which is 1.62 mm.

\end{abstract}

\section{Introduction}


Articulatory-acoustic inversion (A-to-A) aims to recover the vocal tract shape from the acoustic speech signal. This problem has given rise to a large number of approaches, either using explicit acoustic modeling and an articulatory model~\cite{Panchapagesan2011, Ouni2005}, or using database-driven machine learning techniques~\cite{Richmond2015} with recent improvement to incorporate attention mechanisms~\cite{chung24_interspeech}. At present, much of this work is based on electro-magnetic articulography (EMA) available databases~\cite{Wrench2000}.

The technical characteristics of the EMA\cite{Rebernik2021} limit inversion to the front part of the vocal tract. Indeed, EMA requires sensors to be glued on easily accessible articulators. Here, the work focuses on the inversion of the entire vocal tract from the glottis, the complete tongue, the velum to the lips. Here the work relies on a real-time MRI database of more than 3 hours of speech. 

We begin by describing the specific characteristics of rt-MRI imaging compared with EMA, before presenting the data used and the calculation of acoustic features. We then present the method, the loss function used and the evaluation of the inversion. For the experiments we consider the inversion of each articulator separately, or the inversion of all articulators simultaneously. Finally, we present and discuss the results.

\subsection{Specifity of rtMRI with respect to EMA }
The nature of real-time MRI data\cite{Toutios2016} changes the way inversion can be addressed.
The first difference is, of course, that medio-sagittal images cover the entire vocal tract from the glottis to the lips, unlike EMA data (or X‐ray microbeam\cite{Westbury2005}), which correspond to just a few sensors, usually 7 or 8, corresponding to the lips, the lower incisor, 3 or 4 points on the tongue and possibly the velum. These images therefore provide complete information on the position of all articulators, even if for the moment it is only 2D information. This is a decisive advantage, as it is well known that there are compensation mechanisms between articulators that cannot be observed with EMA. In the same way, MRI images provide information on the shape of the whole tongue, and above all it is possible to use the shape of the vocal tract as input for acoustic simulations.

The second difference between EMA data and MRI images is geometric accuracy. That of the EMA data is of the order of 0.3 mm, whereas the pixel size of our images is 1.62 mm in the medio-sagittal plane and the slice thickness is 8\,mm. Even without taking into account post-processing such as contour tracking, the error is necessarily greater than that of EMA. A weakness of MRI is that image acquisition is not instantaneous, taking around 20 ms in the case of the images we use. This means that an articulator such as the lips or tongue can move between the beginning and end of the acquisition, giving a blurred effect and making contour detection difficult. Added to this is the effect of the slice thickness, i.e. 8\,mm, which may be only partially occupied by human tissue, making the distinction between articulators and air difficult. The rt-MRI images are therefore a good approximation of the geometric position of the articulators, but they do not correspond to the physical ground truth, i.e. the actual geometric position of the articulators. 
MRI, on the other hand, requires no gluing of sensors, which has two advantages. The tissues of the tongue or lips to which the sensors are attached are not stiffened by the glue used to fix them, and secondly, no wire passes between the lips, unlike EMA which requires adaptation and slightly disturbs the articulation.
The speech signal that is inverted is the one that is recorded in the MRI machine and then denoised. It is of very good quality (see audio files in~\cite{Isaieva2021}) compared with other MRI databases, but some traces of MRI machine noise remain, whereas the speech signal recorded for EMA data is generally very clean.

The fact that the data are images raises a question about the nature of the inversion results. The first solution is to recover a sequence of images from the speech signal to be inverted~\cite{Csapo2020}. This is the simplest solution, but it is not possible to exploit the results directly, with the drawback of adding inversion errors to the difficulty of interpreting the images due to the above mentioned specificities of MRI. The second solution we have chosen is to use automatically detected articulator contours. We have good-quality contours thanks to the segmentation tools developed based on RCNN~\cite{Ribeiro2021}. The advantage is that we can easily use and evaluate the inversion results.

\section{Dataset}
For this study, we used a dataset containing recordings of one female native French speaker. The dataset was recorded at Centre Hospitalier Régional Universitaire de Nancy and consists of 2,100 sentences, totaling approximately 3.5 hours of recordings with segmentation into 44 phonemes. It contains 153 acquisitions, each lasting 80 seconds and comprising 4,000 images. The rt-MRI images were captured with a resolution of 136 × 136 pixels, a frame rate of 50 fps\cite{Uecker2010}, and a pixel spacing of 1.62 mm. The corresponding audio recordings, sampled at 16 kHz, were recorded using an optical microphone and denoised. The software Astali \cite{fohr2015importance} was used to perform forced alignment of the speech with the transcriptions and to obtain phonetic segmentation. The phonetic annotations were then carefully manually corrected by an expert. 
\subsection{Data preprocessing}
\begin{figure}[ht] 
  \centering
  \includegraphics[width=0.475\textwidth]{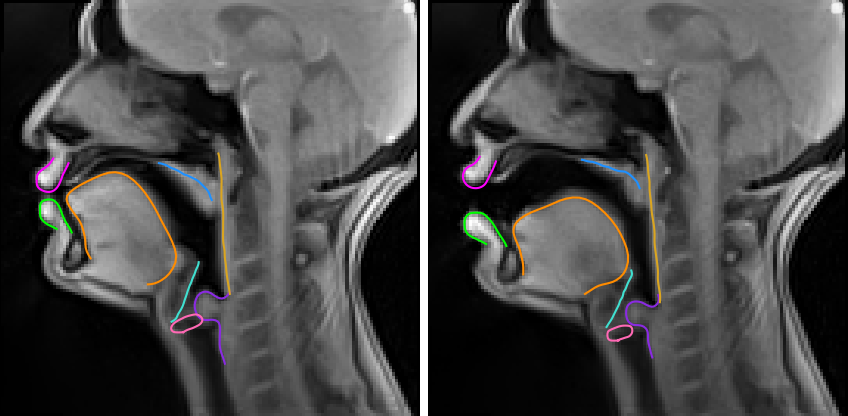} 
  \caption{
    Segmentation of articulators contour tracked in two images of the rt-MRI film:\hspace{0.2cm}
    \raisebox{0.25ex}{\textcolor{arytenoid-cartilage}{\rule{0.2cm}{0.08cm}}} Arytenoid cartilage,
    \raisebox{0.25ex}{\textcolor{epiglottis}{\rule{0.2cm}{0.08cm}}} Epiglottis,
    \raisebox{0.25ex}{\textcolor{lower-lip}{\rule{0.2cm}{0.08cm}}} Lower lip,
    \raisebox{0.25ex}{\textcolor{vocal-folds}{\rule{0.2cm}{0.08cm}}} Vocal folds,\hspace{0.135cm}
    \raisebox{0.25ex}{\textcolor{soft-palate-midline}{\rule{0.2cm}{0.08cm}}} Soft palate midline,\hspace{0.05cm}
    \raisebox{0.25ex}{\textcolor{tongue}{\rule{0.2cm}{0.08cm}}} Tongue,\\
    \raisebox{0.25ex}{\textcolor{upper-lip}{\rule{0.2cm}{0.08cm}}} Upper lip,
    \raisebox{0.25ex}{\textcolor{pharyngeal-wall}{\rule{0.2cm}{0.08cm}}} Pharyngeal wall
  }
  \label{fig:original_contours} 
\end{figure}

We calculated MFCCs along with their first-order ($\Delta$) and second-order ($\Delta\Delta$) derivatives from the audio signal. The number of coefficients was set to 13, with a window size of 25 milliseconds and a hop length of 10 milliseconds. We used an automatic tracking approach based on RCNN ~\cite{Ribeiro2021} to segment the contours of the articulators: Upper lip, Lower lip, Tongue, Soft palate midline, Pharyngeal wall, Epiglottis, Arytenoid cartilage, and Vocal folds (glottis), as shown in Figure \ref{fig:original_contours}. Each articulator contour consists of 50 points with X and Y coordinates.

Phonetic segmentations were encoded in a one-hot format. Using the phonetic segmentations, we removed the silences between sentences, as they do not provide any articulatory information. However, we kept the silences within the sentences for training purposes only, excluding them from the evaluation. This is because we cannot assume that the articulators remain in a neutral position during pauses, as they may involve breathing, swallowing, or other gestures unrelated to speech production. 

MFCCs and articulator contours were normalized following \cite{parrot2020independent}, by subtracting the mean and dividing by the standard deviation. For the contours, the mean and standard deviation were computed using the 50 preceding and following recordings to ensure a consistent distribution of the extracted features. 
Each MFCC frame corresponds to 10 ms, while each MRI image covers 20 ms. To align them, we added an intermediate contour computed as the average of two consecutive contours.

\section{Methods}
\subsection{Model architecture}
\begin{figure}[ht]
    \centering
    \includegraphics[width=0.5\textwidth]{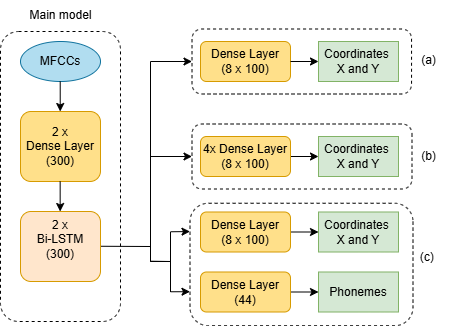} 
    \caption{(a) Single-Task-5, (b) Single-Task-8, (c) Multi-Task-5}
    \label{fig:model_architecture}
\end{figure}

We used two different models, one of which has a variant, as shown in Figure \ref{fig:model_architecture}. The first model and its variant are the same as \cite{azzouz2025complete}. This model consists of five layers: it takes as input a feature vector of dimension 39, passes through two dense layers with 300 units each, followed by two bidirectional LSTM layers, each consisting of 300 units. The output is generated by a dense layer, producing a tensor of size 100 × 8 (number of articulators), where 100 represents the contour points (50 for the X coordinates and 50 for the Y coordinates). We refer to this model as Single-Task-5 (ST-5).

The variant, called multi-task-5 (MT-5), which serves as an alternative to ST-5, provides two outputs instead of one. One output is identical to that of the ST-5 model, while the other is obtained through an additional dense layer that provides classification probabilities for the 44 phonemes of French.

As the ST-5 model was designed to predict only a single articulator, we then add three additional layers with the same output dimension, i.e., 100 × 8 (number of articulators). These additional layers enable the model to classify each point to its specific articulator, improving its ability to handle multiple articulators. We refer to this model as Single-Task-8 (ST-8).

\begin{table*}[b]
\caption{Comparison of RMSE (mm) and MEDIAN (mm) for the ABA and AAT approaches.}
\label{tab:approaches_comparison}
\centering
\begin{tabular}{|l|c|c|c|c|}
\cline{2-5}
\multicolumn{1}{c|}{} & \multicolumn{2}{c|}{\textbf{ABA}} & \multicolumn{2}{c|}{\textbf{AAT}} \\
\cline{2-5}
\multicolumn{1}{c|}{} & \textbf{RMSE} & \textbf{MEDIAN} & \textbf{RMSE} & \textbf{MEDIAN} \\
\hline
\textbf{Arytenoid cartilage} & 1.75 $\pm$\, 1.03 & 1.50 & 1.76$^{*}$ $\pm$\ 1.04 & 1.51 \\
\textbf{Epiglottis}          & 1.68 $\pm$\, 0.94 & 1.48 & 1.69$^{*}$ $\pm$\ 0.96 & 1.48 \\
\textbf{Lower lip}           & 1.69 $\pm$\, 0.93 & 1.49 & 1.73$^{*}$ $\pm$\ 0.94 & 1.52 \\
\textbf{pharyngeal wall}     & 1.25 $\pm$\, 0.67 & 1.11 & 1.21$^{*}$ $\pm$\ 0.64 & 1.08 \\
\textbf{Soft palate midline} & 1.47 $\pm$\, 0.69 & 1.35 & 1.49$^{*}$ $\pm$\ 0.71 & 1.36 \\
\textbf{Tongue}              & 2.42 $\pm$\, 1.19 & 2.18 & 2.58$^{*}$ $\pm$\ 1.25 & 2.33 \\
\textbf{Upper lip}           & 1.35 $\pm$\, 0.65 & 1.21 & 1.37$^{*}$ $\pm$\ 0.67 & 1.23 \\
\textbf{Vocal folds}         & 1.65 $\pm$\, 0.89 & 1.47 & 1.64$^{*}$ $\pm$\ 0.89 & 1.46 \\
\hline
\textbf{Mean}                & 1.65 $\pm$\, 0.87 & 1.47 & 1.69$^{*}$ $\pm$\ 0.99 & 1.49 \\
\hline
\end{tabular}
\caption*{\centering
\small
$^{*}$Significant difference compared to the ABA result ($p < 0.05$) based on a t-test.
}
\end{table*}

\subsection{Loss function}
Acoustic inversion is a regression task in which the Mean Squared Error (MSE) is the most commonly used loss function. Its objective is to minimize the squared differences between the predicted and actual points. 
\begin{equation}
\text{MSE} = \frac{1}{n} \sum_{i=1}^{n} \left( y_i - \hat{y}_i \right)^2\label{eq2}
\end{equation}
where n represents the total number of observations in the dataset, \(y_i\) and \(\hat{y}_i\) represent the true and predicted values of the output for example \(i\), respectively.

When the task also involves classification, such as phoneme recognition, the Cross-Entropy loss function is used to measure the discrepancy between the predicted probability distribution of each phoneme and the actual phoneme labels.
\begin{equation}
\text{Cross Entropy} = - \sum_{i=1}^{n} \sum_{c=1}^{C} y_{i,c} \log(\hat{y}_{i,c})\label{eq1}
\end{equation} 
\(C\) is the total number of classes, \(y_{i,c}\) and \(\hat{y}_{i,c}\) represent the true and predicted probabilities, respectively, that example \(i\) belongs to class \(c\).

To predict contours while simultaneously classifying phonemes, both loss functions are combined.
\begin{equation}
L(y_i, \hat{y}_i) = \text{MSE}(y_i, \hat{y}_i) + \text{CrossEntropy}(y_i, \hat{y}_i)\label{eq3}
\end{equation}
\subsection{Evaluation of the model}
We evaluated the unnormalized root mean square error (RMSE) and the median, both expressed in millimeters. For each frame, we computed the RMSE for each articulator, which means the RMSE between the 100 predicted and original points, corresponding to the geometry of each articulator. Then, we calculated the RMSE and the median average across all frames for each articulator, and then average them across all articulators.
\begin{equation}
\text{RMSE} = \sqrt{\frac{1}{n} \sum_{i=1}^{n} \left( y_i - \hat{y}_i \right)^2}\label{eq4}
\end{equation}

\begin{table*}[ht]
\caption{Comparison of RMSE (mm) and MEDIAN (mm) for the AAT approach across the ST-5, ST-8, MT-5 and ST-5-cw11 models}
\label{tab:model_comparison}
\centering
\begin{tabular}{|l|c|c|c|c|c|c|c|c|}
\cline{2-9}
\multicolumn{1}{c|}{} & \multicolumn{2}{c|}{\textbf{ST-5}} & \multicolumn{2}{c|}{\textbf{ST-8}} & \multicolumn{2}{c|}{\textbf{MT-5}} & \multicolumn{2}{c|}{\textbf{ST-5-cw11}}\\
\cline{2-9}
\multicolumn{1}{c|}{} & \textbf{RMSE} & \textbf{MEDIAN} & \textbf{RMSE} & \textbf{MEDIAN} & \textbf{RMSE} & \textbf{MEDIAN} & \textbf{RMSE} & \textbf{MEDIAN}\\
\hline
\textbf{Arytenoid cartilage} & 1.76 $\pm$\ 1.04 & 1.51 & 1.76 $\pm$\ 1.09 & 1.49 & 1.78$^{*}$ $\pm$\ 1.05 & 1.53 & 1.74$^{*}$ $\pm$\ 1.04 & 1.49 \\
\textbf{Epiglottis}          & 1.69 $\pm$\ 0.96 & 1.48 & 1.69 $\pm$\ 0.96 & 1.49 & 1.69 $\pm$\ 0.95 & 1.48 & 1.66$^{*}$ $\pm$\ 0.93 & 1.46 \\
\textbf{Lower lip}           & 1.73 $\pm$\ 0.94 & 1.52 & 1.76$^{*}$ $\pm$\ 0.97 & 1.55 & 1.67$^{*}$ $\pm$\ 0.92 & 1.46 & 1.74$^{*}$ $\pm$\ 0.95 & 1.53 \\
\textbf{pharyngeal wall}             & 1.21 $\pm$\ 0.64 & 1.08 & 1.22$^{*}$ $\pm$\ 0.65 & 1.09 & 1.22$^{*}$ $\pm$\ 0.65 & 1.09 & 1.20$^{*}$ $\pm$\ 0.64 & 1.07 \\
\textbf{Soft palate midline} & 1.49 $\pm$\ 0.71 & 1.36 & 1.49 $\pm$\ 0.71 & 1.36 & 1.46$^{*}$ $\pm$\ 0.69 & 1.33 & 1.46$^{*}$ $\pm$\ 0.70 & 1.33 \\
\textbf{Tongue}              & 2.58 $\pm$\ 1.25 & 2.33 & 2.67$^{*}$ $\pm$\ 1.29 & 2.40 & 2.50$^{*}$ $\pm$\ 1.19 & 2.27 & 2.57$^{*}$ $\pm$\ 1.24 & 2.32 \\
\textbf{Upper lip}           & 1.37 $\pm$\ 0.67 & 1.23 & 1.38$^{*}$ $\pm$\ 0.69 & 1.24 & 1.36$^{*}$ $\pm$\ 0.66 & 1.22 & 1.35$^{*}$ $\pm$\ 0.65 & 1.22 \\
\textbf{Vocal folds}         & 1.64 $\pm$\ 0.89 & 1.46 & 1.66$^{*}$ $\pm$\ 0.92 & 1.46 & 1.68$^{*}$ $\pm$\ 0.90 & 1.50 & 1.62$^{*}$ $\pm$\ 0.88 & 1.43 \\
\hline
\textbf{Mean}                & 1.69 $\pm$\ 0.99 & 1.49 & 1.70$^{*}$ $\pm$\ 1.02 & 1.47 & 1.67$^{*}$ $\pm$\ 0.96 & 1.46 & 1.67$^{*}$ $\pm$\ 0.98 & 1.45 \\
\hline
\end{tabular}
\caption*{\centering
\small
$^{*}$Significant difference compared to the ST-5 result ($p < 0.05$) based on a t-test.
}
\end{table*}

\subsection{Experiments}

In this study, we conducted two types of experiments to evaluate articulator prediction. The first approach, called articulator-by-articulator (ABA), involves training a separate model for each articulator. The second approach, referred to as all-articulators-together (AAT), aims to predict all articulators simultaneously. To compare these two approaches, we used the ST-5 model.

Next, for the second experiment we explored different experimental configurations exclusively within the AAT approach. We compared several model architectures, including ST-5 and ST-8. We also tested MT-5, which is capable of predicting both articulatory contours and phonetic segmentation. This comparison allowed us to assess the impact of phonetic segmentation accuracy on the model’s performance.
Additionally, we evaluated the impact of the context window by testing a configuration with a context window of 11 frames, consisting of 5 previous frames, 5 subsequent frames, and the current frame, to enrich temporal information, as \cite{parrot2020independent}, to investigate the role of temporal information in the model’s performance. We refer to this model as Single-Task-5-cw11 (ST-5-cw11).

\subsection{Model parameters}
All models were trained for 500 epochs with a batch size of 10, using the Adam optimizer with an initial learning rate of 0.001. Early stopping was applied with a patience of 10 epochs on the validation data, halting training if no improvement was observed. Our dataset contains 450,000 images after removing silence.We randomly divided our dataset by acquisitions into 80\% for training, 10\% for validation, and 10\% for testing. All experiments were run with the same training configuration and train-validation-test splits to ensure a fair comparison. The entire implementation was done using PyTorch.

\section{Results}

Table \ref{tab:approaches_comparison} compares the performance of two approaches: articulator-by-articulator (ABA) and all-articulators-together (AAT). We evaluate these methods using both root mean square error (RMSE) and median error (MEDIAN), expressed in millimeters. All models converged before 200 epochs.
Overall, the ABA approach achieves a slightly lower mean RMSE of 1.65 mm compared to 1.69 mm for AAT. When analyzing individual articulators, ABA performs better for most, particularly for the Arytenoid cartilage (1.75 mm), Epiglottis (1.68 mm), Lower lip (1.69 mm), Soft palate midline (1.47 mm), Tongue (2.42 mm), and Upper lip (1.35 mm). However, AAT shows slight improvements for certain articulators, such as the Pharyngeal wall (1.21 mm) and Vocal folds (1.64 mm). Notably, the Tongue exhibits a difference of 2.58 mm for AAT versus 2.42 mm for ABA.For the median error, values remain close across both methods, with an average of 1.47 mm for ABA and 1.49 mm for AAT. This suggests that error distributions are similar despite differences in RMSE. Finally, the comparison between ABA and AAT reveals nearly identical overall performance.

Table \ref{tab:model_comparison} presents the RMSE and median values for all-articulators-together (AAT) approach across four models: ST-5, ST-8, MT-5, and ST-5-cw11. For the Arytenoid cartilage, RMSE values ranged from 1.74 mm to 1.78 mm, with median values between 1.49 mm and 1.53 mm. The Epiglottis showed consistent performance, with RMSE between 1.66 mm and 1.69 mm and median values from 1.46 mm to 1.49 mm. The Lower lip had RMSE values between 1.67 mm and 1.76 mm and median values from 1.46 mm to 1.55 mm. For the Pharyngeal wall, RMSE ranged from 1.2 mm to 1.22 mm, with median values between 1.07 mm and 1.09 mm. The Soft palate midline showed minimal variation, with RMSE between 1.46 mm and 1.49 mm and median values from 1.33 mm to 1.36 mm. The tongue had the highest RMSE values, ranging from 2.5 to 2.67 mm, while the median values ranged from 2.27 to 2.4 mm. The Upper lip showed similar performance across all models, with RMSE between 1.35 mm and 1.38 mm and median values from 1.22 mm to 1.24 mm. The vocal folds had RMSE values between 1.62 mm and 1.68 mm, with medians ranging from 1.43 mm to 1.5 mm.

The ST-5-cw11 model achieved the lowest RMSE and median values across all articulators, except for the lower lip and the tongue. It was followed by the MT-5 model, which obtained the lowest RMSE and median values for three articulators: the lower lip, the tongue, and the soft palate midline which is similar to ST-5-cw11. The ST-5 model ranked third, while ST-8 had the worst scores in most cases.

The mean RMSE was lowest for ST-5-cw11 and MT-5 (1.67 mm), followed by ST-5 (1.69 mm), with the highest value for ST-8 (1.7 mm). The median error was also lowest for ST-5-cw11 (1.45 mm), followed by MT-5 (1.46 mm), ST-8 (1.47 mm), and the highest value for ST-5 (1.49 mm).

\section{Discussion}
The results show that the ABA approach, which achieved an RMSE of 1.65 mm, slightly outperforms the AAT approach,  which have an RMSE of 1.69 mm. This suggests that training each articulator individually yields better results. However, in practical applications, it would be more relevant to use a single model capable of predicting all articulators. 

Since contour tracking is not perfect and may introduce errors during training, we have incorporated phonetic segmentation as additional labels. It is important to note that the phonetic segmentation was manually verified in its entirety, ensuring that it does not introduce any errors. This additional information aims to compensate for tracking inaccuracies. However, as shown in Table  \ref{tab:model_comparison}, the difference between training with and without phonetic segmentation is minimal.

When comparing the different models used for AAT, we observe that MT-5 and ST-5-cw11, which includes a context window of 11, both achieved an RMSE of 1.67 mm. This suggests that integrating phonetic segmentations and increasing the context window size enhance performance.

On the other hand, the deepest model, ST-8, obtained the worst score, with an RMSE of 1.70 mm, confirming that our baseline model, ST-5, is sufficient for this prediction task. Although these differences remain minimal, they provide valuable insights for optimizing model architectures.

\begin{figure}[ht] 
  \centering
  \includegraphics[width=0.475\textwidth]{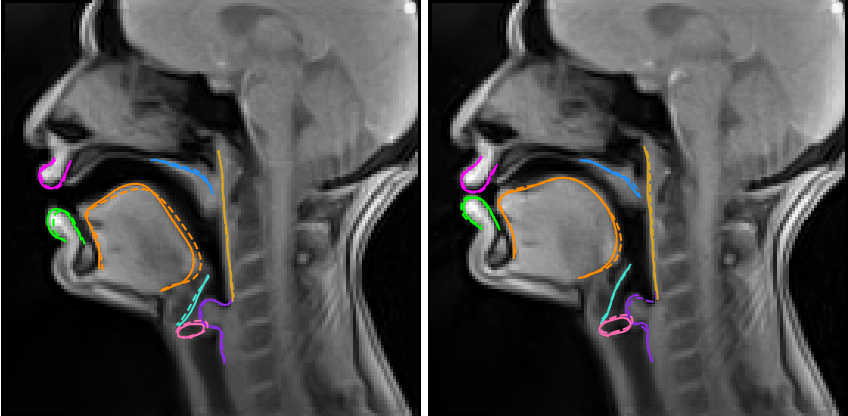}
 \caption{
    Example of two inversions compared to the original contours tracked in the rt-MRI film.  
    The dotted lines show the predicted contours, while the solid lines represent the original ones.  
    The RMSE is 1.65 mm for the phoneme 'p\_cl' on the left and 1.46 mm for 'E/' on the right.
}
  \label{fig:predicted_contours} 
\end{figure}

Furthermore, while the quality of contour segmentation is high, it remains imperfect, and the model is inherently limited by the accuracy of the provided segmentation. Compared to previous studies, our results are reasonably strong, as shown in Figure \ref{fig:predicted_contours}. However, our approach is distinctive in its ability to predict the complete vocal tract contour, which is not achieved by existing methods.


\section{Conclusion}

In this work, we have demonstrated the possibility of inverting the entire vocal tract. The presented acoustic-to-articulatory model exploits denoised audio signal data and contours extracted from MRI images. The results show an average error (RMSE) of 1.65 mm for the ABA approach and 1.69 mm for the AAT approach to be compared to the pixel size of 1.62 mm.

Even though the average RMSE precision is satisfactory, some errors remain. One alternative would be to integrate articulatory variables, which offer a different perspective from simple geometric distance and provide a more relevant evaluation from a phonetic point of view or to associate the recovery of the vocal tract contours to the reconstruction of an inverse image via diffusion models\cite{nguyen2024speech2rtmrispeechguideddiffusionmodel}.

Another alternative would be to use the natural speech recorded outside the MRI machine. Even though the denoised audio signal is of good quality, it was recorded in a noisy environment in a supine position. We thus plan to use the normal speech signal to have an inversion that better matches a potential application and to be able to adapt to any speaker.
\clearpage
\bibliographystyle{IEEEtran}
\bibliography{mybib}

\end{document}